\documentclass[a4paper,fleqn]{cas-dc}

\usepackage[numbers]{natbib}

\def\tsc#1{\csdef{#1}{\textsc{\lowercase{#1}}\xspace}}
\tsc{WGM}
\tsc{QE}

\begin{document}
\let\WriteBookmarks\relax
\def\floatpagepagefraction{1}
\def\textpagefraction{.001}

\shorttitle{A greedy strategy to attract more girls to CS}    

\shortauthors{T. Catarci, L. Podo, D. Raffini and P. Velardi}  

\title [mode = title]{Programming Skills are Not Enough: a Greedy Strategy to Attract More Girls to Study Computer Science}  

\tnotemark[<tnote number>] 


%

\author[1]{Tiziana Catarci}

\cormark[1]


\ead{<catarci@diag.uniroma1.it>}


\credit{Research design, paper writing}

\affiliation[1]{organization={Sapienza Università di Roma},
            addressline={piazzale Aldo Moro, 5}, 
            city={Roma},
            postcode={00100}, 
            country={Italy}}

\author[1]{Luca Podo}


\ead{podo@di.uniroma1.it}
\credit{Implementation of questionnaire, data collection, graphical design, software}


\author[1]{Daniel Raffini}


\ead{raffini@diag.uniroma1.it}
\credit{Literary review, methodology and project description}


\author[1]{Paola Velardi}
\cormark[1]

\ead{velardi@di.uniroma1.it}
\credit{Research design, design and analysis of follow-up program, paper writing}


\cortext[1]{Corresponding authors}



\begin{abstract}
This paper addresses the issue of the gender gap in ICT, which hinders the digital transition of many countries. We present a methodology (along with two use cases) that leverages girls' greater sensitivity towards social issues - such as sustainability and health - to mitigate some well-known stereotypes that make ICT disciplines, and software engineering particular, unattractive to girls.
Our strategy is based on a “grid” approach which aims to strengthen both some important vertical skills (e.g., environmental sustainability and programming) and horizontal ones, such as public speaking, team-building and team-working, the ability to compete, and social networking. As remarked by virtually all studies concerned with the design of effective teaching strategies, one of the main points to consider is that teaching technical skills is not enough. The importance of soft skills in education is underestimated, and there is an urgency to make up for the lack of these skills. Furthermore, many studies have noted that the lack of soft skills affects girls more than boys.

\end{abstract}


\begin{highlights}
\item We provide a literature review to analyze some of the main causes of the lack of women professionals in software engineering, some of which are underlooked in state-of-the-art studies on the gender gap in ICT; 
\item We build on previous projects to combat the gender gap in ICT - which often include mentoring programs, the use of role models, and gaming activities - by proposing a teaching methodology aimed both at highlighting the role of ICT technologies in addressing societal issues, and at strengthening soft skills.   
\item We present a strategy, that we call ''greed(y)" because it has been conceived as a \textit{grid} of vertical (hard) and horizontal (soft) skills, intertwining topics to which girls are traditionally sensitive, such as environmental sustainability, health, etc., with digital skills and soft skills that the education system more rarely considers - such as team-working, public speaking, social networking, and competition -  with visible consequences more for girls than for boys. 
\item We describe two projects, directed at high-school girls, university students, and young professionals,  aiming to demonstrate the effectiveness of the greedy strategy when applied to real cases. In this regard, we note that ''greedy" is also a term used in Computer Science to denote sub-optimal strategies, as in our case, since we acknowledge that, in order to achieve a higher impact, similar programs should be proposed much earlier in a student's career. Nevertheless, the ongoing digital transformation and the absence of workers with
adequate skills in Computer Science requires short-term actions.
\end{highlights}


\maketitle

\section{Introduction and Motivations}\label{intro}

Among STEM disciplines, ICT is the one with the highest gender unbalance, with the percentage of women enrolled in graduate programs usually ranging between 10 and 20 percent depending on the country. During the last decade, the significant and long-lasting gender gap in the field of Computer Science has been pointed out by many studies (\cite {Saxa17} and \cite{jaccheri2021gender}, among others). The gender gap exists at all levels: university students, researchers, professors, public administration, and industry. A cascading effect is created due to a lack of awareness and action at the lowest levels of education. As a result, digital transformation is currently predominantly guided by men, and this does not guarantee the diversity of positions and ways of thinking that would be indispensable in such an important field.

The shortage of female ICT professionals contributes to the severe shortage of skilled digital workers that is retarding many countries’ development. The lack of digital skills affects companies and public administrations at different levels: there are not enough experts who can develop cutting-edge technology and employees who are able to use existing technology to improve efficiency. 

In recent years, the topic has also entered the agenda of policymakers. Next Generation EU (also known as the Recovery Fund) includes digitization, innovation, competitiveness, culture, green revolution, and ecological transformation among the most funded objectives of national plans. However, the risk is that those benefiting from the funds made available by the European Community are mainly male-dominated sectors (such are indeed digital, construction, agriculture, and transport), even though the pandemic has had devastating effects primarily on female employment. Therefore, the need for actions to reduce the gender gap in these areas is extremely urgent. 
Since 2014 the European Commission has been monitoring the digital progress in the Member States by means of the Digital Economy and Society Index (DESI3), which analyses four key areas: Human capital, Connectivity, Integration of digital technology, and Digital public services. The European Commission also launched the Digital Skills and Jobs Platform. The DESI also focuses attention on the gender gap, considered one of the causes of the lack of digital workers: from the 2022 DESI Report\footnote{\url{https://digital-strategy.ec.europa.eu/en/library/digital-economy-and-society-index-desi-2022}} emerges that only 19 percent of ICT specialists are female.

As surveyed in \cite{jaccheri2021gender}, numerous projects have been proposed by public and private stakeholders to combat the gender gap in ICT.  Among the most adopted strategies, the authors cite mentoring programs, exploitation of female role models, and the use of gaming activities to arouse a greater passion for coding.
While we agree on the effectiveness of these strategies, we believe that educational projects must also contain other elements aimed both at highlighting the role of ICT technologies in addressing societal issues, and at enhancing soft skills.

In this paper, we present a methodology (along with two use cases) that leverages girls' greater sensitivity towards social issues - such as sustainability and health - to mitigate some well-known stereotypes that make ICT disciplines, and software engineering particularly, unattractive to girls.  We call our methodology ''greedy" both because it is conceived as a grid of technical (vertical) and soft (horizontal) skills, and because it is sub-optimal, being aimed at girls in high schools or universities, while we are well aware that similar programs should start much earlier in a student's career.

We contribute to the state-of-the-art literature on educational programs to combat the gender gap in ICT in many ways:
\begin{enumerate}
\item we propose to exploit the greater tendency of women to choose careers that have social implications, by emphasizing the central role of ICT technologies in mitigating many societal issues; 
\item we enrich our educational program by focusing on the enhancement of soft skills, the absence of which in educational programs penalizes girls more than boys;
\item we propose a novel strategy, that we name greedy, conceived as a \textit{grid} of vertical (hard) and horizontal (soft) skills;
\item we present two practical examples of applications of this strategy;
\item we propose a follow-up program to verify in the long and short range the effects of our strategy, and we present the first results of this program. To the best of our knowledge, this is the first time a follow-up strategy to monitor the effectiveness of educational programs on the gender gap in ICT has been proposed and discussed. 
\end{enumerate}
The paper is organized into a literary review, a methodology section, and two case studies. The literary review (Section \ref{literature}) underlines the importance of soft skills in software engineering education and shows how this is related to the gender gap issue. Section \ref{method}  explains the greedy strategy we propose as a theoretical framework to better organize projects to retain girls into ICT careers. The last two sections are focused on the description of two projects held at the Sapienza University of Rome: the WomENcourage 2019 conference (Section \ref{womencourage}) and the G4GRETA project (Section \ref{g4greta}), both aiming at demonstrating the effectiveness of the greedy strategy when applied to real cases. Regarding the G4GRETA project,  follow-up results are offered in Section \ref{followup}.  Finally  Section \ref{conclusions} summarizes the main findings and provides suggestions for future actions. 

\section{Literature review}\label{literature}

Even though Computer Science is the area where there are the most job offerings and the highest salaries, girls still do not enroll in ICT degree programs and do not pursue careers in Computer Science. The factor that drives girls away from Computer Science disciplines is the persistence of gender stereotypes and biases, which lead people to consider computer science-related jobs as inappropriate for women. Unlike other types of prejudices, the one relating to women’s limited aptitude for information technology appears rooted and more difficult to weaken, because it tends to condition girls’ interests and inclinations from an early age, leading to a rejection of information technology disciplines which is very difficult to break down in adolescence or adulthood.

While boys, when choosing a course of study, are particularly attracted by the possibilities of career and economic progression, for girls social and innovation impact are a very relevant factor. In \cite{Leonhardt21} the authors show that for boys the main attraction in choosing to study Computer Science is a passion for video games, while girls like computer games less. In \cite{Berg} it is demonstrated how a strong deterrent for girls is the stereotype of a nerd, as an antisocial individual with poor communication skills, thus antithetical to their aspirations. In \cite{Cheryan} it is suggested that the media is largely responsible for the perpetuation of this kind of stereotype. It is no coincidence that Artificial Intelligence, whose fascinating applications have been widely covered by the media in recent times, attracts girls instead, as \cite{Vachosky} demonstrates. In corroboration, the bachelor’s degree program in Applied Computer Science and Artificial Intelligence started in 2020/2021 by the Department of Computer Science at Sapienza University of Rome has already reached 35 percent of girls’ enrolment. Similarly, the course in Philosophy and Artificial Intelligence achieved gender equality.

We may conclude that the gender gap in ICT is not the consequence of difficulty for women in accessing job positions (although some obstacles remain to reach the top positions), but rather a cultural resistance of the girls themselves when choosing their course of study. Compared to other fields, where women face objective difficulties in advancing their careers, in the case of Computer Science it is the discipline itself that does not attract school-age girls, because of deeply rooted stereotypes at the societal level. To mitigate this problem and redirect girls into the Computer Science field, many initiatives have been launched at the European and national levels. In Italy has recently been created the portal GICT (Atlas of Gender Initiatives in ICT)\footnote{\url{https://www.gict.it/}}  to collect and connect the numerous national initiatives, while at the European level is to be mentioned EUGAIN (European Network for Gender Balance in Informatics)\footnote{\url{https://eugain.eu/}}, whose aim is to improve gender balance at all levels through an international network. 

Many of the past and current initiatives have been aimed at female high school students. Some of these projects achieved some results on a small scale but failed to reverse the general trend. The reason is explained in \cite{Master}, where the authors have pointed out that undermining the prejudices of girls with respect to Computer Science is very difficult in adolescence, suggesting that effective awareness programs on computer disciplines should be offered in pre-school or lower school age, that is the moment when gender biases take root in girls’ mind. Nevertheless, even if the most effective way to proceed is to intervene at a younger age, the ongoing digital transformation and the absence of workers with adequate skills in Computer Science requires fast action. There is a need to keep acting on the high school age range, even if the bias is already deep-rooted in girls, by looking for effective compensatory measures to show girls the potential of the Computer Science field. 
For the reason given, it is important to design effective strategies and methodologies that consider the complexity of the problem and face it efficiently. 

 As remarked by virtually all studies concerned with the design of effective teaching strategies in the domain of software engineering, one of the main points to consider is that teaching technical skills is not enough.  Many studies have pointed out the importance of soft skills in software engineering education and agree that there is an urgency to make up for the lack of soft skills which affects girls more than boys when relating to Computer Science. Soft skills are required from industries and are considered fundamental abilities for a software engineer. In \cite{Oguz} the authors try to map out the gaps between academia and industry, in relation to soft skills, mention: being fluent in English, teamwork, problem-solving, critical thinking, and communication. In \cite{Bona23} the gap in the software engineering learning process from the perspective of IT team leaders is investigated, in order to make the teaching closer to the industry request. Among the results, the authors cite shortage of communication, management skills, patience, resilience, and pro-activity, all of which can be attributed to a lack of attention in the teaching of soft skills. In \cite{Alves} the authors propose a Project-Based Learning methodology to mitigate the scarcity of preparation of software engineering graduates by involving students in projects required by a ''real" customer. The study concludes that the development of soft skills is one of the results acquired by students facing real projects.

In \cite{Sedelmaier}, a body of skills for software engineering is proposed. It is based on data-driven research design, which includes non-technical skills in addition to technical ones. Soft skills are important since developing software requires teamwork, and the capability to understand and communicate with stakeholders. For this reason, the paper suggests that more attention should be paid to the educational aspects of software engineering. The aim is to produce competencies, that arise from the interaction of hard and soft skills embedded in a real, emotional context taking into account ethics and regulations. A software engineer needs context-sensitive skills (e.g., communication skills related to a specific field), generic soft skills (e.g., presentation skills), and factual knowledge (the specific subject domain). In \cite{Sedelmaier} a method that implies interviews with successful software engineers is proposed, also defining a framework composed of some macro-groups of skills: collaboration, communication, ability for structuring, self-reflection, consciousness of problems, competence to solve problems, respect of ethical and social norms, responsibility, writing. 

Another work \cite{Groeneveld}  offers a systematic literary review of papers addressing the issue of teaching soft skills in software engineering programs. The methodology, in this case, consists of searching the most frequently mentioned soft skills in software engineering education papers, and identifying self-reflection, conflict resolution, communication, and teamwork as the top taught skills. The authors also identify levels at which to integrate soft skills (lectures of a single course, project within a course, throughout the entire curriculum, capstone projects, internships) some teaching aspects (interdisciplinarity, group composition, collaborative tools, assessment tools, active learning, video watching, playful learning). 

Although soft skills are often underlooked in schools, boys have more opportunities to develop them. In \cite{Marcenaro} it is demonstrated that soft skills are shaped by socioeconomic conditions and familial contexts. 
One of the ways through which these skills can be honed is participation in sports. Boys, more than girls, tend to be involved in competitive and team sports. These activities require constant communication, strategy planning, and mutual support among teammates, all of which contribute to the development of soft skills. They learn how to handle victory and defeat, understand the importance of resilience, and develop leadership qualities. In \cite{Siddiky} the author examines the variation of soft skill development according to co-curricular activities. The study shows that males develop soft skills more than females except one which is presentation skills.

On the flip side, societal biases and gender stereotypes play a significant role in hindering the development of soft skills, particularly among women. From a young age, girls might be dissuaded from participating in activities deemed 'masculine,' such as competitive sports, which inadvertently limits their exposure to situations where soft skills are developed. The lack of predisposition to competition in women as a result of a gender stereotype has been demonstrated by \cite{Sutter}. Due to prevailing biases, women are often less encouraged to speak in public or take on leadership roles. This lack of encouragement stems from stereotypical notions about gender and leadership, where assertiveness and decisiveness —traits often associated with leadership — are attributed exclusively to men.

In \cite{Groeneveld} creativity is also mentioned  as a skill for software engineers, while the authors in \cite{Seibel} demonstrate that re-establishing the idea of software engineering as a creative job is one of the keys to attracting women and eradicating self-prejudice. A further relationship is thus settled between the development of soft skills and the strategies for bringing women into Computer Science. In \cite{Seibel} the authors analyze the relationship between women and CS education by interviewing women at Coding Bootcamps, which are attended mainly by adult women who have already faced the working environment. It results that those exposed to coding before college are more likely to choose a CS degree. Many women realized that maths is not their main skill, but logical reasoning is, and feel more confident working in women's groups or groups where women are equally represented. In \cite{Seibel}, the authors also believe that it is important to communicate the wide spectrum of applications of software engineering. 

Based on this literature review, we can conclude that two undervalued aspects when designing an educational program to attract more girls to computer science studies are the need to combat the nerd stereotype that makes ICT disciplines unattractive for girls, and to pay more attention to soft skills.

\section{Methodology}\label{method}

In  \cite{Sedelmaier} three areas of research are proposed concerning software engineering pedagogy: identification and description of competencies, experiments with didactic approaches, and an assessment framework to evaluate the effectiveness of the adopted approaches. From a methodological point of view, these areas can be considered as the study of a theory, a field of experimentation, and a follow-up study, which are the three stages of the strategy that we describe hereafter. 

We propose a new methodology, that we call “greedy”, aimed at attracting girls from high school and encouraging them to enroll in ICT degrees. The methodology has been already tested on two projects held at Sapienza University in Rome: the 2019 edition of the ACM  celebration of Women in Computing  (hereafter WomENcourage 2019) and the G4GRETA Project. Furthermore, we present the first results of a follow-up program conducted on G4GRETA students, to measure in a more systematic way the effectiveness of the proposed approach. We remark that very rarely systematic follow-up programs have been presented in the literature. 

The term  ''greedy"  refers to two aspects of the method shaped in this paper:
\begin{enumerate}
\item A strategy conceived as a grid of vertical (hard) and horizontal (soft) skills, intertwining topics to which girls are traditionally sensitive - such as environmental sustainability, health, and creativity - with digital skills and soft skills that the public education system more rarely considers - such as team-working, public speaking, social networking, and competition - with visible consequences more for girls than for boys (see Section \ref{literature}).
\item A sub-optimal strategy, since we acknowledge that, in order to achieve a higher impact, similar programs should be proposed much earlier in a student’s career. As noted previously, however, the scarcity of professional profiles in the software engineering sector, and ICT in general, requires urgent interventions. The term refers to an approach to solving a problem by selecting the best option currently available, similar to greedy algorithms.
\end{enumerate}
As we discussed, misconceptions concerning ICT-related professions, such as the “nerd” stereotype and the belief that programming consists only of typing weird symbols, affect more females than males. Computer Science curricula in schools are, depending on the country, absent, deficient, or late, in the sense that they are introduced when these prejudices are well-established. Likewise, education systems often lack programs aimed at cultivating horizontal or soft skills, which are known to be very important in career development. The progress of women in any career is hampered not only by a male-dominated organization of society but also by the fact that girls’ education in families and schools does not favor the learning of essential soft skills, such as the ability to compete, public speaking, teamwork, and networking. Outside of school life, boys have the opportunity to acquire some of these skills by practicing more masculine sports, which could justify the results of a recent study in which a substantial gap was observed between the performance of men and women in teamwork and team competition.

In conclusion, girls have fewer opportunities to develop both soft and hard skills. To mitigate these deficiencies, our  method is based on two principles:
\begin{itemize}
     
\item To attract girls’ interest in ICTs at a later stage of education, it is necessary to make them understand that Computer Science has significantly contributed to the advancement of other disciplines or subjects in which women traditionally show great passion and aptitude, such as medicine, environment and sustainability, art, and social sciences.

\item Furthermore, it is necessary to integrate technical skills with horizontal skills, essential for progressing in the professional career, especially in male-dominated sectors such as ICT.
\end{itemize}
For these reasons, we have conceived a grid-shaped approach in which Computer Science is one of the vertical (hard) skills that are taught and intersected with other vertical and horizontal (soft) skills, such as those previously listed. We believe that empowering women is a complex process that requires not only upgrading their skills in strategic and traditionally male-dominated sectors, such as ICT and software engineering in particular, but also acquiring soft skills that provide them with the right tools to progress in their careers. This can help girls understand that computational skills nowadays are very important in different careers and are becoming an increasingly required competence to access all levels and fields of the job market. It is also important to make girls understand that ICT disciplines open up many professions that are not limited to being programmers but can offer a very strong bond with society, culture, economy, and business.

The context in which the greedy strategy proved to be most effective in raising awareness
of equal opportunities concerned two academic projects, both managed by two ICT Departments of Sapienza University of Roma: 
\begin{enumerate}
\item The first one, the 2019 edition of the ACM Celebration of Women in Computing (WomENcourage 2019), brought together women in the computing
profession and related technical fields, to exchange knowledge and experience,
and to provide special support for young women who were pursuing their academic degrees or starting their careers in computing. Through a program packed with insightful
interdisciplinary topics and engaging educational and networking activities,
WomENcourage provided a unique experience of collective energy, leadership,
and excellence that professional women must share to support each other. 

\item The other program is G4GRETA (Girls for Green Technology Applications),
which involves third and fourth-year students from high schools in and around
Rome. This  project - which started in November 2022 and is now in its second edition - combines
the development of hard and soft Computer Science skills with the theme of
Environmental Sustainability. The participants learn to design apps with an eco-sustainable
theme, with the help of tutors and university lecturers. G4GRETA proposes
lessons, exercises, and training courses that foster the self-determination of
young female students in technical-scientific career paths with a holistic approach,
interconnected to environmental sustainability, encouraging female students to develop
digital applications and solutions that embrace innovation.
\end{enumerate} 
In the following Sections, we provide additional details on these projects.

\section{Project 1: WomENcourage 2019}\label{womencourage}
\subsection{Description}\label{womendescription}

WomENcourge is an annual conference organized  under the patronage of the Association for Computing Machinery
whose aim is to bring together women in the computing profession and related
technical fields to exchange knowledge and experience and provide special support for
women who are pursuing their academic degrees and starting their careers in computing.
This conference has reached in 2023 its tenth edition and has a well-assessed program
including a hackathon, a career fair, workshops, and keynotes. The 2019 edition\footnote{\url{https://womencourage.acm.org/2019/}}, although starting from this initial program, has been enriched with activities and
events according to the grid scheme previously described.

As regards the objective of strengthening vertical (hard) skills, we have
combined workshops on traditional ICT topics such as gaming, cybersecurity, data science, software engineering
and more, with several events concerning interdisciplinary topics that notoriously
attract women, such as: artificial intelligence and health; smart cities; architecture, neuroscience and technology for conscious cities; art and artificial intelligence. Interdisciplinary topics
were given higher visibility, organizing them in the main Auditorium, in the form of
Mini-conferences, keynotes, and Panels, with a focus on the scientific quality of all the
invited speakers, who were selected to inspire the young attendees.

The major innovation was in the number and quality of activities aimed at strengthening
vertical (soft) skills, among which the hackathon, the poster presentations, and the
poster karaoke. The latter was a rather challenging activity in which girls were asked to
present each other’s poster, assigned by drawing among the 10 best posters selected at
the end of the standard poster sessions. This activity was aimed at fostering mutual listening,
the ability to improvise, and to face challenges even when one feels unprepared. In fact, robust evidence exists that women tend to have lower self-esteem and attitude to risk than men \cite{selfesteem}, being much less trained than men to increase these skills. 

\subsection{Summary and results}\label{womenresults}

\begin{figure*}
  \centering
  \includegraphics[width=\linewidth]{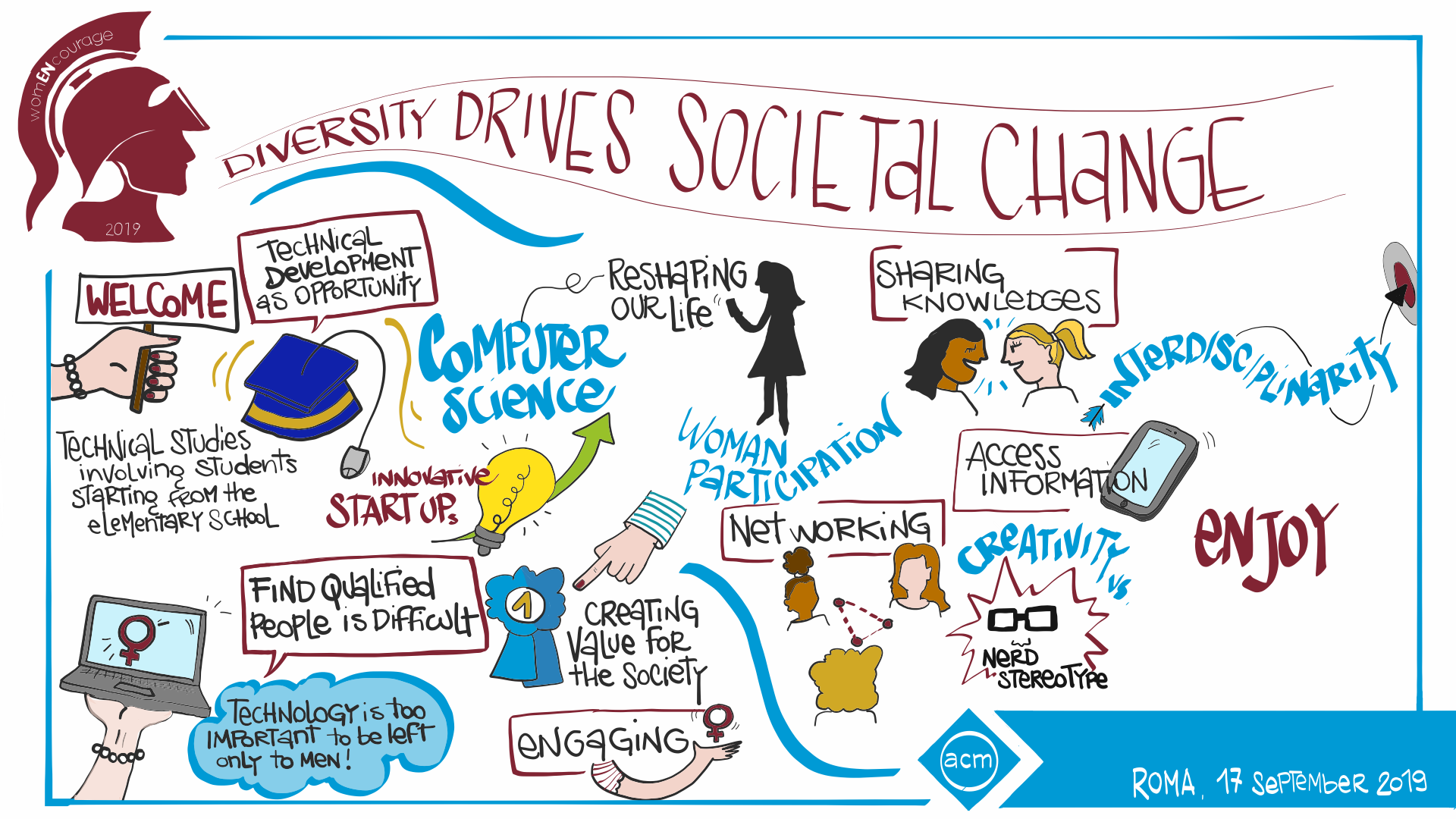}
  \caption{One amongst the many infographics generated in real-time during the WomENcourage 2019 event, synthesizing the relevant keywords of the event.}
  \label{fig:infographic}    
\end{figure*}

The sixth edition of WomENcourage took place in Rome, at Museum MAXXI, from September 16th to 19th, 2019. More than 350 participants took part in the event, 55\% of whom were young: 62\% of them graduates, and 38\% still university students. WomENcourage also showed a good presence of male participants, who represented 15\% of all participants (primarily involved in the organization of some events as speakers, company representatives during the career fair, etc.).  The participants came from 32 countries and covered 4 different continents (Europe, Asia, Africa, and America). In summary, the rich program included a hackathon, a career fair, one laboratory, 9 workshops, 2 mini-conferences, 2 panels, 3 poster sessions, a poster karaoke, two scientific and four technical keynotes, and a number of recreational events. To inspire creativity and fun, a designer created and projected the infographics of all the main events in real-time (see Figure \ref{fig:infographic}). 

To date, WomENcourage 2019 remains the conference edition with the largest number
of participants, and the largest number of sponsoring and/or supporting partners
(both national and international) to demonstrate the great interest not only of the girls
involved but also of companies, public bodies, and non-profit organizations.

\section{Project 2: G4GRETA}\label{g4greta}
\subsection{Description}\label{g4gretadescription}
WomENcourage 2019 was the first experiment in applying our greedy strategy,  within the existing framework of the ACM Celebrations of Women in Computing. We were able to see how the girls, after some initial resistance, showed themselves to be very open to the challenges represented by the activities aimed at strengthening soft skills, and extremely interested in the interdisciplinary themes proposed in the conference. This experimentation represented the basis on which to build a long-term and more complex project,  not aimed at girls who had already undertaken studies or professions related to ICT, but at high school students who have not yet chosen a course of study, G4GRETA. 

The G4GRETA (Girls for GREen Technology Applications) project intercepts many elements of the Next generation EU \footnote{\url{https://www.europarl.europa.eu/RegData/etudes/BRIE/2020/652000/EPRS_BRI(2020)652000_EN.pdf}}  program:
digitization and innovation, the green revolution, and the gender gap, albeit
on a small scale, such as a pilot program on digital and environmental
culture, aimed at female students of high schools in Lazio, Italy.

The objectives of this initiative were manifold. The most obvious is to encourage greater access for women to leadership roles, including technological ones, which today are almost exclusively male-dominated. In turn, this could increase the push towards a green evolution of the current consumer society, in which women usually demonstrate greater sensitivity towards social and environmental problems. In this regard, an article published in December 2017 by Scientific America titled “Men Resist Green Behavior as Unmanly” states the following: ”Women have long surpassed men
in the arena of environmental action; across all age groups and in all countries, women
tend to lead a greener lifestyle. Compared to men, women throw less waste, recycle
more and leave a smaller carbon footprint. Some researchers have suggested that personality
differences, such as women’s altruism priority, may help explain this gender
gap in ecological behavior”. Whatever the explanation, it seems clear that greater involvement of women in technology choices can only advance the green economy more
rapidly.

\begin{figure*}
  \centering
  \includegraphics[width=\linewidth]{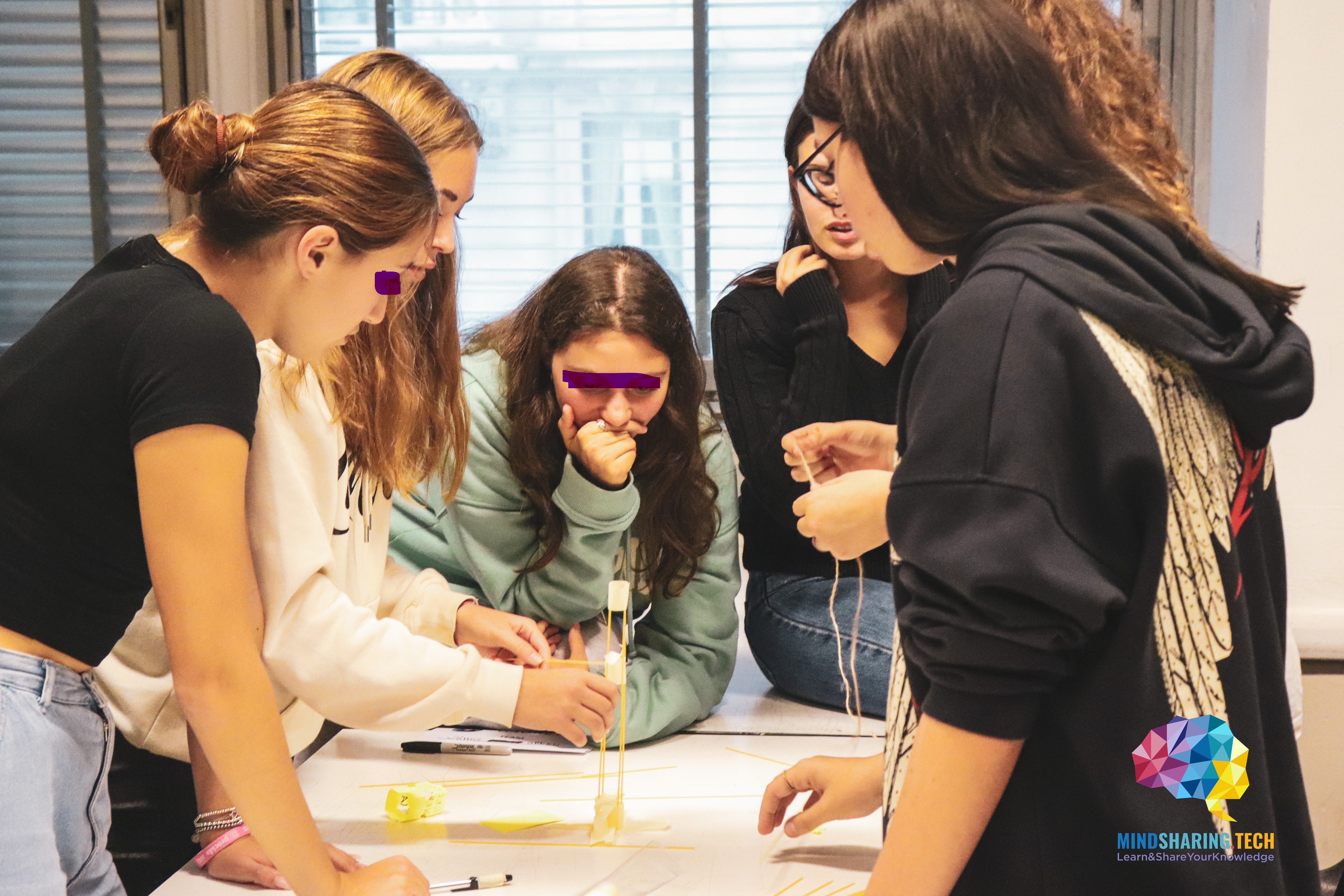}
  \caption{A team engaged in a teambuilding game during the first edition of G4GRETA}
  \label{fig:teamgame}    
\end{figure*}

Following our "greedy" strategy, the project is organized into 10 meetings, during which girls increase their awareness of environmental issues, learn how ICT technologies can support the ecological transition, and are provided with the basic principles of coding and problem-solving.  In addition, they attend dedicated teambuilding  (see Figure \ref{fig:teamgame}), public speaking, video making, and social networking meetings. The teaching material is available on the project website, also for replicability purposes. 
At the end of the program,  participants are asked to submit: 1) the project of a “green application”; 2) a description of the considered problem and its solution, using the design thinking approach; 3) a short video, to illustrate their work\footnote{For the 23-24 edition, the video will be inspired to the "dance your PhD" contest \url{https://en.wikipedia.org/wiki/Dance_Your_Ph.D.}, to create more engagement and fun.}.  A kick-off meeting is dedicated to illustrating in detail the objectives, deadlines, and methodology to carry on their team projects. During this meeting, teams are encouraged to adopt the design thinking methodology (which was the object of a dedicated lesson) to support their creative process.

Next, for about 45 days, the teams develop their projects autonomously and remotely, with the support of tutors (university students) and mentors. A first selection of the best projects is performed by a team of judges selected among university professors and the external partners of the project (private and public bodies). The proponent teams are then asked to illustrate their projects in a public session. Finally, the 3 best projects are prized during the final meeting, held in some “inspiring” location (on May 2023, the selected location was the Botanical Garden of Roma). 

The first edition of the project started on November 22, and the new edition (23-24) is about to start at the time we are writing.  Overall, each edition 
 spans over a period of 7 months, followed by a timeframe  during which we follow the cohort of female students involved
in the program, to evaluate its impact. For its entire duration, the project also includes
a community building and management program, through social media, some of which
(such as Linkedin) aimed at connecting girls  with professionals and role models,
others (Tiktok, Instagram) are more suitable for networking among participants and other peers.

The proposed initiative has involved a large number of beneficiaries:
\begin{enumerate}
    \item School teachers: they are made aware of ICT and green issues, and then independently
- or with advice from the University - they can design similar educational projects
for their students;

\item High school students: they acquire technical (horizontal) skills in the field of
software design and green technologies, as well as soft skills. Furthermore, they can create a solid network of friendships and relationships with companies, role models, and university students, thanks to dedicated social channels and specific events organized during the follow-up period.






\item Students, doctoral students, and young researchers of the University: they have
been involved in the tutoring of school students, increasing their awareness
of gender issues, being encouraged to network with each other and with the
participating students, strengthening their sense of community, and reducing the discomfort of working in predominantly male contexts;

\item University professors: they are involved in the design and implementation of the
initiative, acting as lecturers, role models, and mentors, thus contributing to an important
goal such as that of gender equality in ICT. They also have the opportunity
to strengthen relationships with the schools, companies, and bodies participating in
the initiative;

\item Companies, institutions, and local bodies: the project also aimed at involving public
and private institutions. We obtained the active participation of IBM Italy, the
Foundation Mondo Digitale (FMD), and the Lazio Region - that co-financed the
initiative in 22-23 - and Ernst \& Young (EY) Foundation, the main partner of the 23-24 edition. All of these organizations are already engaged on the issues of gender
equality and the green transition. Participation in the G4GRETA project will
therefore make it possible to develop priority issues for these organizations and
consolidate relationships with schools and universities.
\end{enumerate}

Twenty-four schools have joined the first edition of the
project, of which 15 are in the city of Rome and 9 are in the Lazio Region.   More than
600 candidates applied, but only 216 were eventually selected due to limitations of
space and resources. 30 school teachers were actively following the initiative. Similar numbers apply to the new edition which is about to start at the time we are writing.

During the meetings the attention of the girls has always remained high, likely as a consequence of the great effort made by the organizers to keep it high, promoting interactivity,  minimizing the time dedicated to teaching lessons, and challenging them with funny but not so simple tasks. We observed with surprise that, after an initial phase of stress, the girls reacted very effectively to our requests to get out of their comfort zone, answering questions in public, engaging in activities never performed before (coding, pitching), and working in a team with other girls they had never met before rather than, as they would have preferred, with their schoolmates. This demonstrates, if proof were needed, that disinterest in information technology, or resistance to public exposure and competition, are certainly not female characteristics, but only a limitation of the educational system, which does not engage them sufficiently in these activities.

\subsection{Follow-up program}\label{followup}

With a view to evaluating the impact of these projects, we considered it very important to design follow-up actions, in order to identify the real effect of the programs on female students when choosing to enroll in a university degree.
This is even more important if we consider that the major part of the projects aimed at filling the gender gap in ICT do not implement, to the best of our knowledge,  follow-up studies to monitor their effectiveness over time.  Follow-up strategies are essential to help identify best practices that need to be implemented at scale, and to inform similar projects, in order to impact the number of women enrolled in Computer Science degree programs. This leads us to a theme of great importance, which is the necessity to identify effective strategies and, subsequently, to apply them at the national and European levels.

Follow-up programs are usually designed in the medical area, and are more difficult to apply to social science because they need continuous funding and a long-term activity. The challenge is the lack, in the current literature, of follow-up studies that may constitute examples for high school orientation projects, although, for a minority of projects, success factors such as an increase of women enrollments have been reported \cite{jaccheri2021gender}. Nevertheless, we consider that a follow-up methodology is a crucial part of a project like G4GRETA. It serves as a bridge to understanding the effectiveness of an intervention and the long-term impacts of the initiative. 

Our follow-up methodology has been conceived as a structured way to maintain communication with participants after the end of the project, to monitor their behavior and choices. 
It consists of three main elements:
\begin{enumerate}
    \item Delivering a survey to understand how girls’
perceptions have changed in relation to the world of Computer Science, as a result of the greedy strategy. 
\item Planning regular in-presence events and continuous social engagement through the project's social channels,  to keep all the project participants
connected and active.
\item Use the most enthusiastic participants of each edition as supporters for subsequent editions.
\end{enumerate}

The ultimate goal of the follow-up program is to monitor how many girls will eventually choose a
university program in Computer Science.  Since the choice of a university program is expected, in the case
of the G4GRETA girls, in two years, it is necessary to maintain contact and team up
with the girls until that time. 
The follow-up strategy will help identify best practices that need to be implemented
on large scale, and similar projects to actually impact enrollment numbers in Computer
Science degree programs. This leads us to a second theme of great importance, which
is the necessity to identify effective strategies and, subsequently, to apply them at the
national and European levels.

\subsubsection{Survey Results}
 
The purpose of the survey is to understand if the self-perception of girls in relation to ICT,  programming, and other soft and hard skills has changed, as a result of the greedy strategy. 
hereafter, we present our results using Sankey diagrams, to facilitate a visual understanding of the connections amongst the different questions in the survey. This type of diagram shows the mass flow from each response type to the subsequent one,  proportional to the flow quantity.

  \begin{figure}
  \centering
  \includegraphics[width=\linewidth]{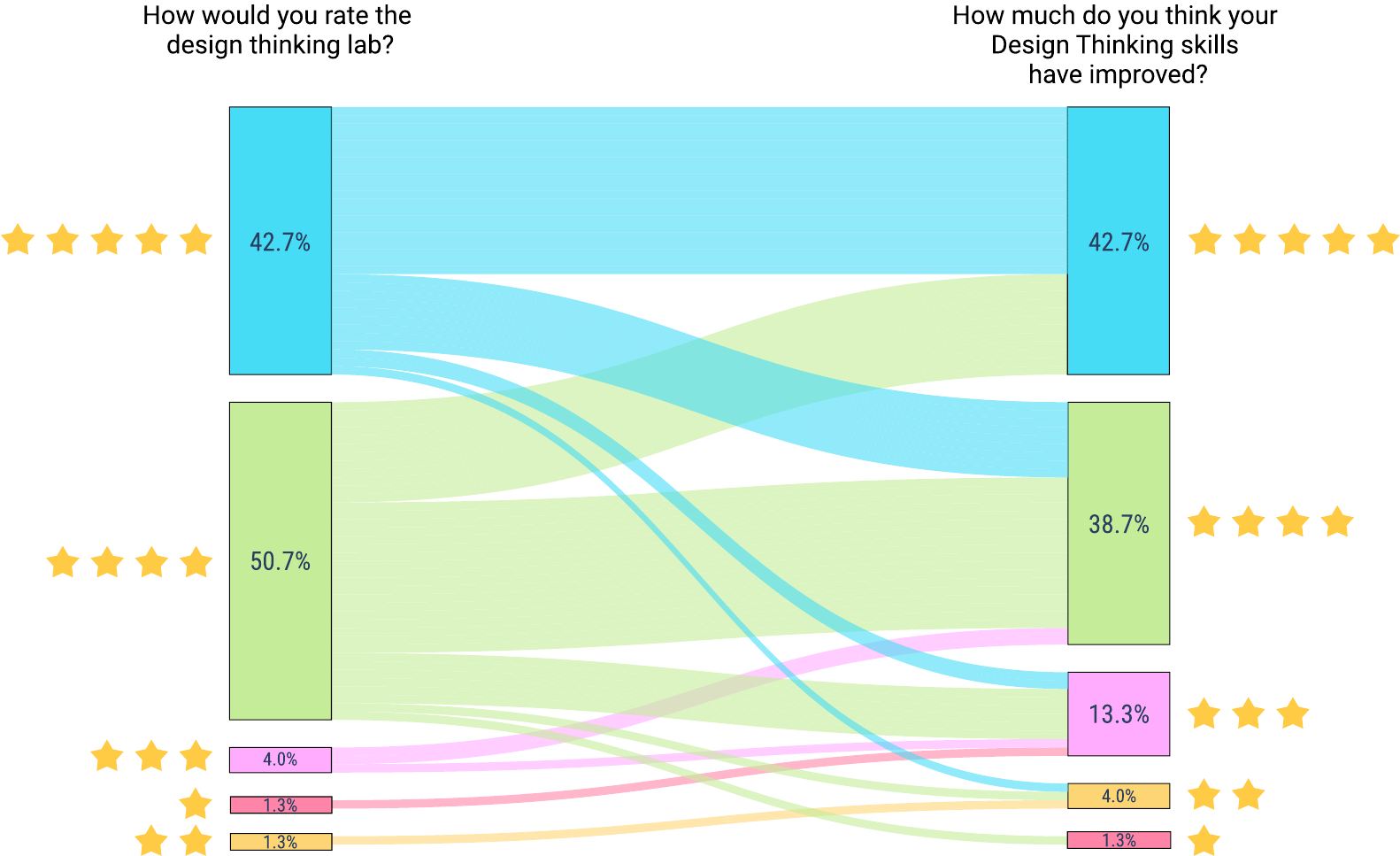}
  \caption{Evaluation of the design thinking activity  and self-perception of improvement}
  \label{fig:sankeydesignthinking}    
\end{figure}

\begin{figure}
  \centering
  \includegraphics[width=\linewidth]{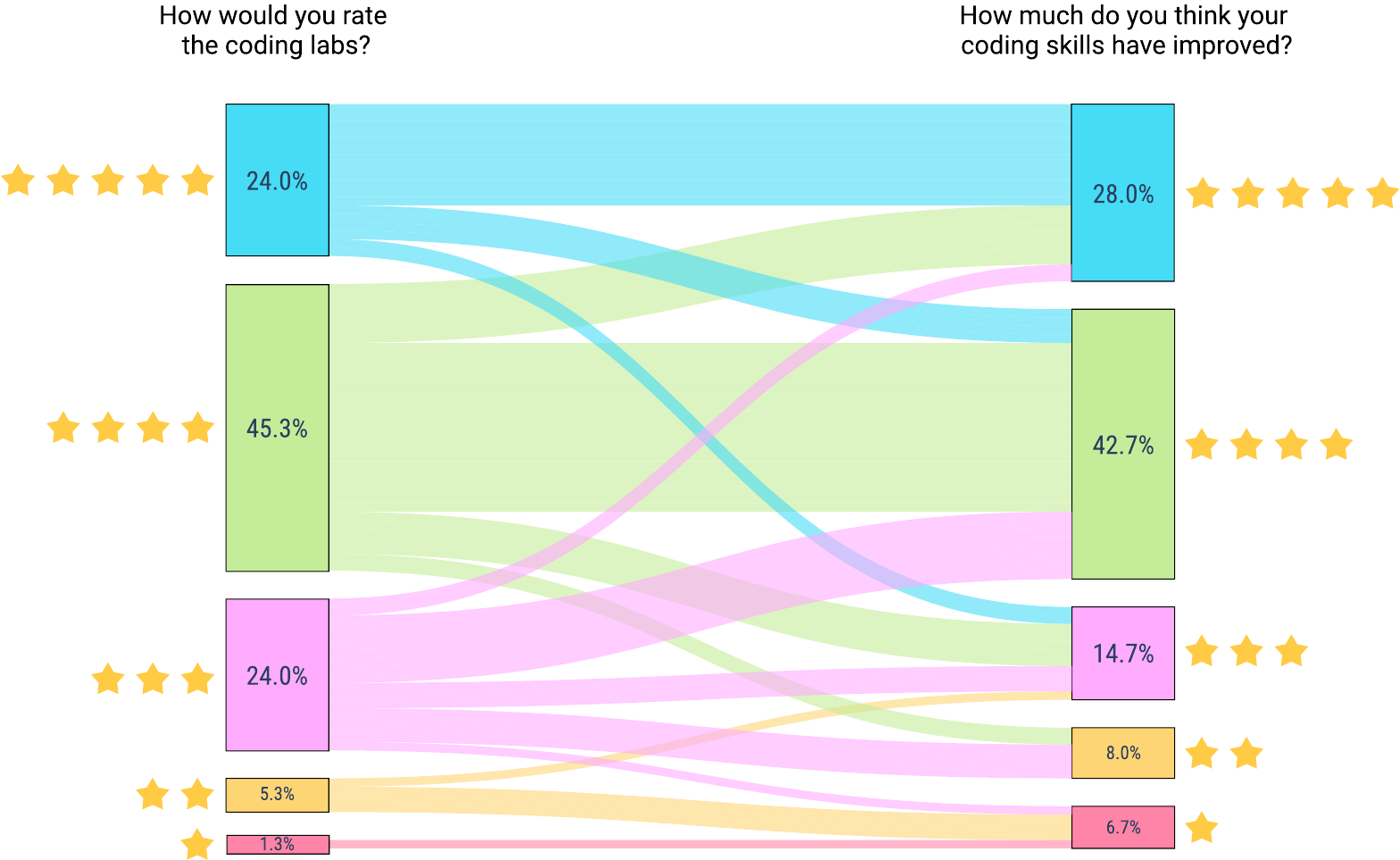}
  \caption{Evaluation of the coding activity  and self-perception of improvement}
  \label{fig:sankeycoding}    
\end{figure}


\begin{figure}
  \centering
  \includegraphics[width=\linewidth]{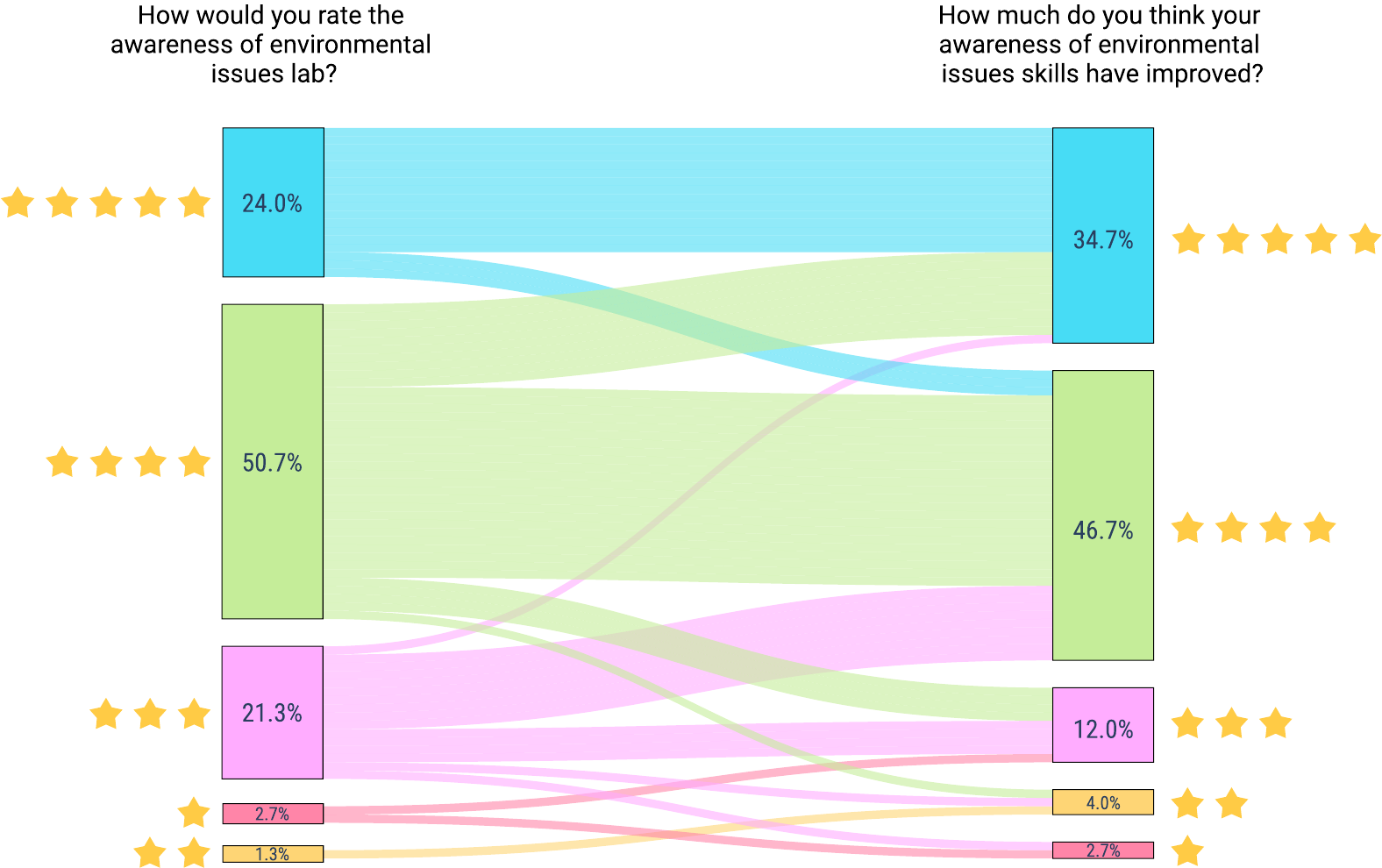}
  \caption{Evaluation of the lesson on technology for the environment   and perceived improvement in awareness}
  \label{fig:sankeyenv}    
\end{figure}

\begin{figure}
  \centering
  \includegraphics[width=\linewidth]{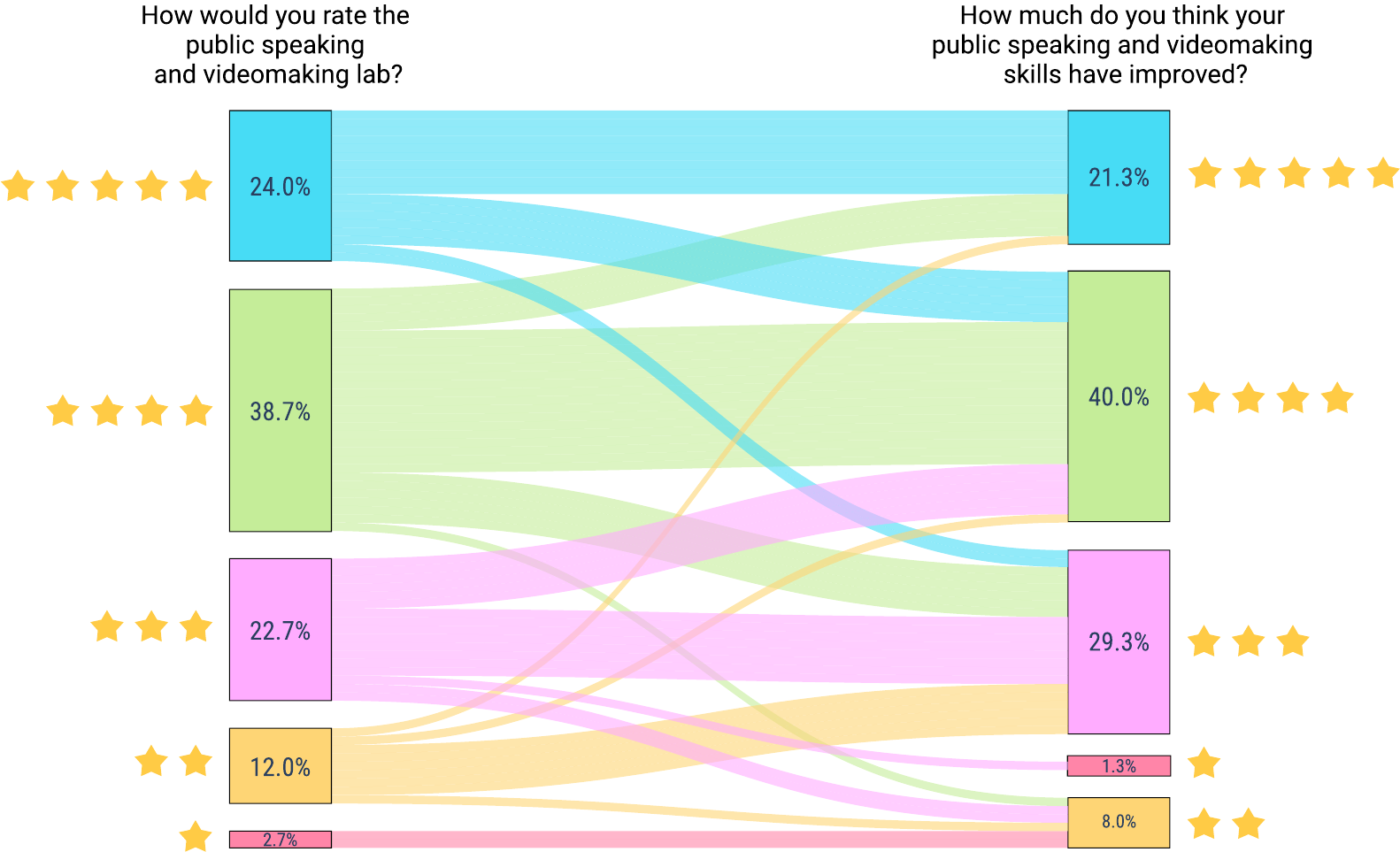}
  \caption{Evaluation of the public speaking activity  and self-perception of improvement}
  \label{fig:sankeypublicspeaking}    
\end{figure}

\begin{figure}
  \centering
  \includegraphics[width=\linewidth]{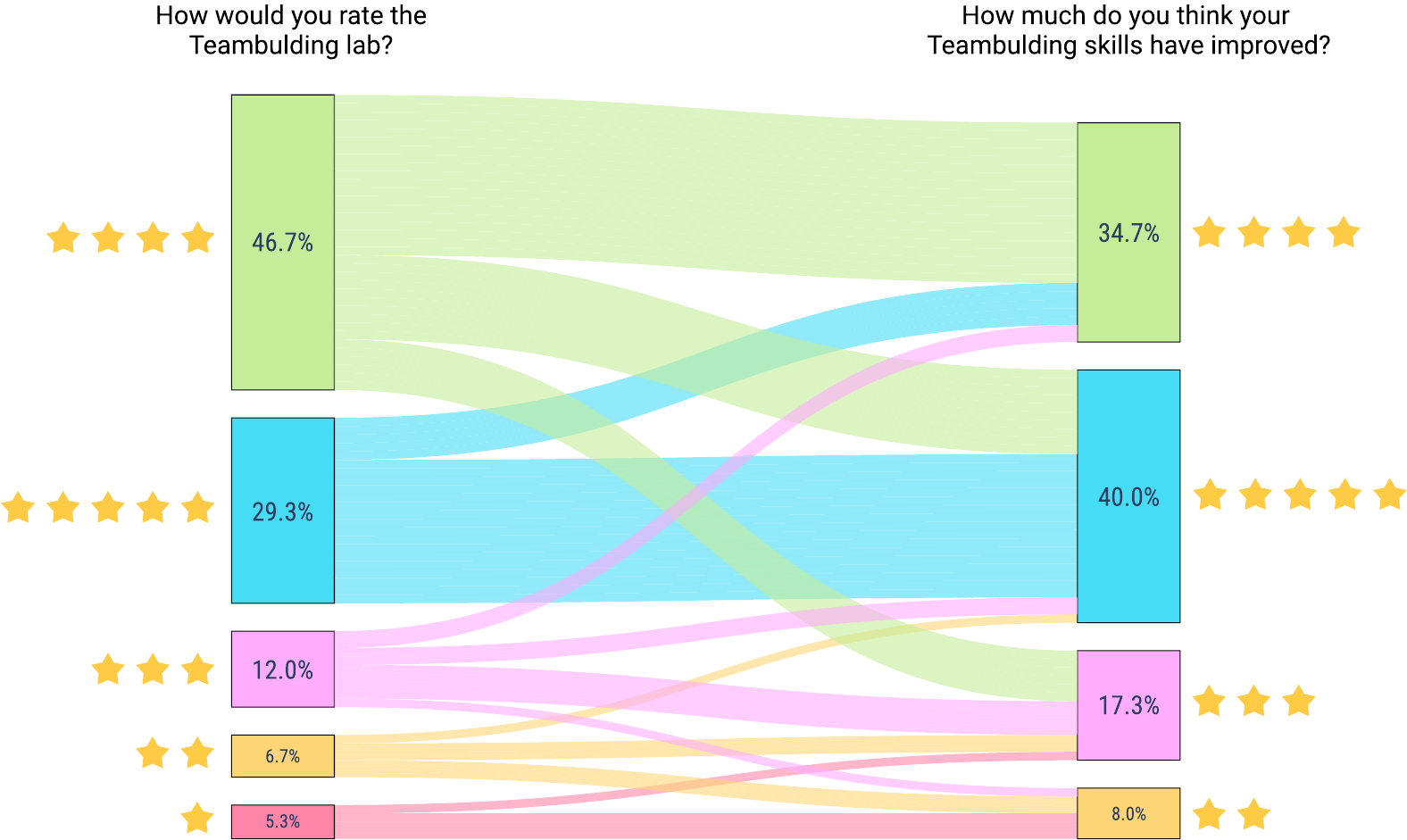}
  \caption{Evaluation of the teambuilding activity  and self-perception of improvement}
  \label{fig:sankeyteambuilding}    
\end{figure}
\begin{figure}
  \centering
  \includegraphics[width=\linewidth]{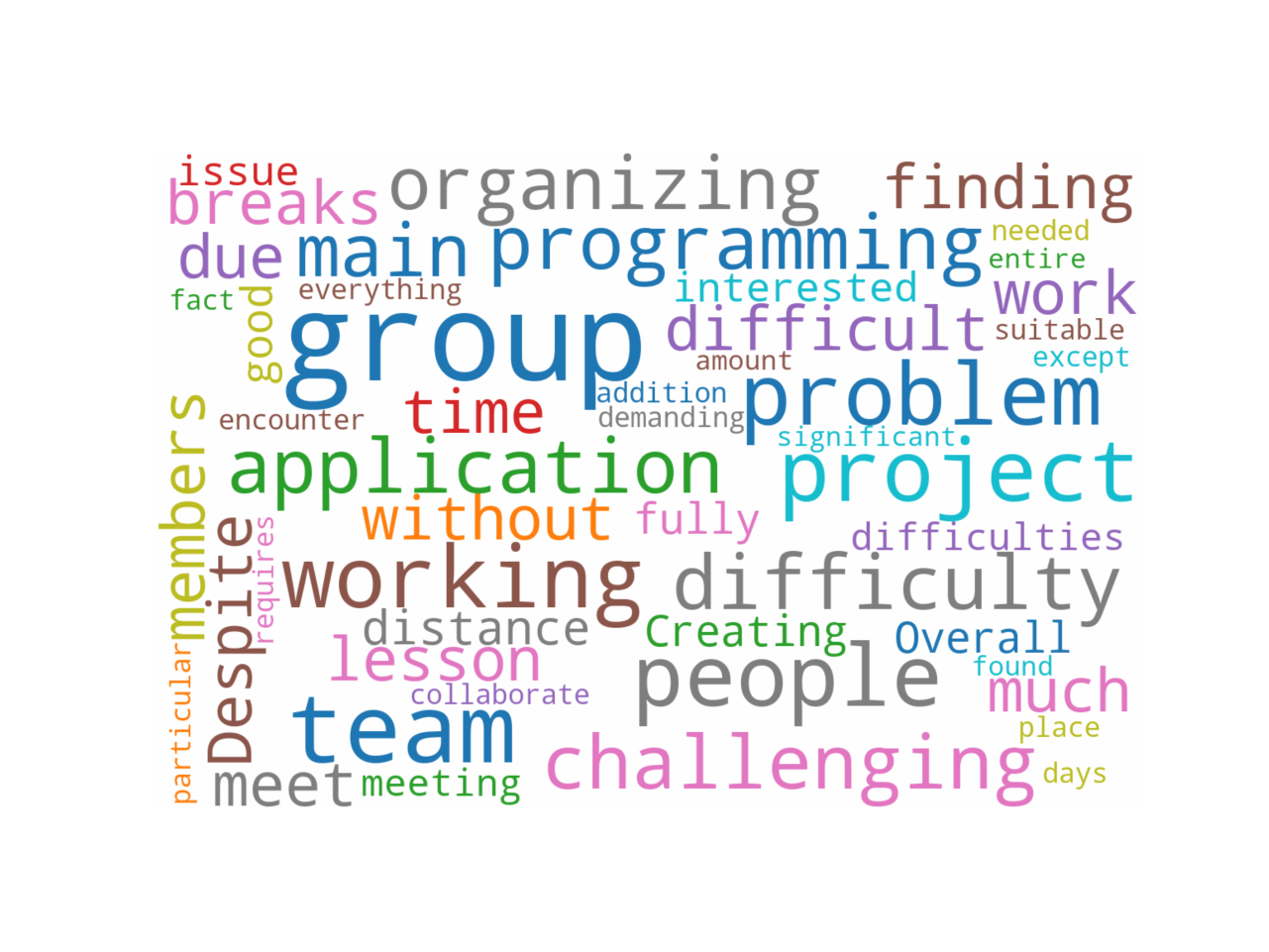}
  \caption{Word cloud describing girls' major difficulties during the project}
  \label{fig:wordcloud}    
\end{figure}
Figures \ref{fig:sankeydesignthinking}, \ref{fig:sankeycoding}, \ref{fig:sankeyenv}, \ref{fig:sankeypublicspeaking}  and \ref{fig:sankeyteambuilding}
 show how much each specific activity has been appreciated by the participant and how much girls' self-perception concerning the related skill has changed after the project. First, we note that all activities have been rated with 4 or 5 stars, including programming. Similarly, all girls perceive that their skills have improved on all topics. It does not come as a surprise that design thinking was the most highly rated activity since this laboratory was centered on a persona-based approach that favors empathy. In this regard, we note that, although the literature suggests that women are more emphatic than men, whether this ability reflects a  difference between women and men or a gender-role stereotype remains an open question, as suggested in \cite{emphaty}. Finally, we note that team building is the skill that girls feel they have improved on the most, as shown in Figure \ref{fig:sankeyteambuilding}.
\begin{figure*}
  \centering
  \includegraphics[width=\linewidth]{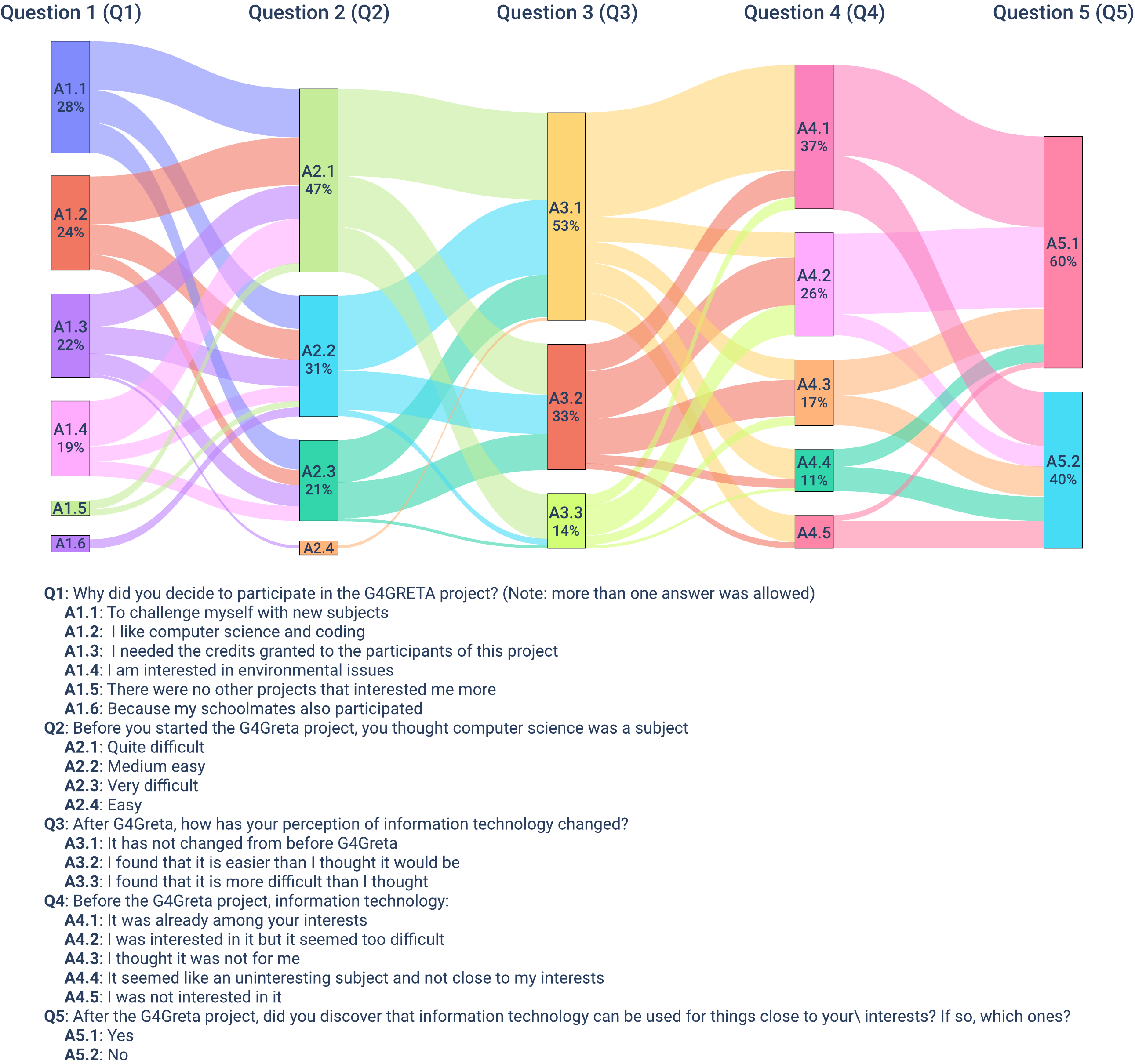}
  \caption{Sankey Diagram of the Follow-up Questionnaire: motivations and impact}
  \label{fig:sankeytotal}    
\end{figure*}

 Figure \ref{fig:wordcloud}  summarizes the main difficulties encountered by the participants. Although, as expected, programming the apps and presenting their projects to the jury caused some stress, we noted that words like ''group", ''team", and ''organizing" have great relevance. This mirrors what was just said about teamwork. If, on the one hand,  girls certainly understood and appreciated the importance of teamwork,  they also realized the difficulty of working in a group. During subsequent interviews, many girls complained that the difficulty in cooperating with other teammates was one of the major sources of stress during the development of individual projects. These considerations will inform better strategies during the next edition of the project.  

 The most interesting results come from the analysis of the Sankey diagram in Figure \ref{fig:sankeytotal}. In particular, answers to question 2 show that almost all the participants believed that Computer Science was quite difficult, very difficult, or medium easy, with a negligible percentage of ''easy". However, 33\% of the respondents eventually changed their opinion to ''I found it is easier than I thought it would be".  We also note that, in the end, 60\% of the girls changed their minds,  discovering that information technology can be used for things close to their interests.  
 Overall, our results in relation to our main objective (raising the interest of girls in programming and computer science) cannot be defined as striking, but certainly satisfactory when considering what was stressed at the beginning of this paper:  the prejudice regarding women's interest in ICT is rooted in nursery and primary school, and much more timely action should be taken.

\subsubsection{Engagement strategy and use of peer models}
Although the first edition of the project ended a few months ago\footnote{With respect to the time we are writing this report}, several follow-up actions have been already undertaken to keep the participants connected and engaged. We list here the main actions and events:
\begin{enumerate}
\item The winning team was interviewed, with the aim of creating a video that could illustrate the project to future participants through the experiences of peers. The video is showcased on the home page of the project website.
\item Some participants and one of the organizers were able to talk about the project as part of the Italian Green broadcast, a program on the national video channel RAI2 dedicated to environmental sustainability.
\item The project was showcased during the October 2023 Maker Faire in Rome, and around 20 girls from the project took turns participating in some activities and acting as testimonials.
\item With the support of the main partner of the 23-24 edition, EY Foundation, we are outlining a social engagement project, which also includes a visit by the winning teams to the company headquarters and a mentoring program.
\end{enumerate}


\section{Discussion and Concluding Remarks}\label{conclusions}

The two initiatives that have been put forth serve as compelling exemplifications of the importance of incorporating soft skills education into orientation programs specifically tailored to guiding young girls toward the field of Computer Science. Soft skills, encompassing a wide array of abilities such as effective communication, collaborative teamwork, critical and analytical thinking, as well as creative problem-solving, are indispensable in today’s fast-paced and ever-evolving technological landscape. These skills not only enhance the girls’ capability to adapt to and excel in various professional scenarios but also play a pivotal role in boosting their confidence and resilience in facing the unique challenges posed by the tech industry. Moreover, the projects underscore the critical necessity of introducing follow-up mechanisms as an integral part of these orientation programs. Such mechanisms are imperative for providing a comprehensive and ongoing evaluation of the initiatives’ impact and efficacy. They enable educators, mentors, and program facilitators to track the participants' progress, assess the development of their soft skills, and gauge the extent to which these programs have succeeded in instilling a genuine interest and proficiency in Computer Science.

The projects that have been presented show promising results, but it is crucial to also highlight some limitations that emerge from the analysis, with the goal of refining and strengthening the adopted methodological approach. Alongside the limitations, we propose some mitigation measures. The first limitation concerns the age of intervention. As suggested by the name (greedy) of our teaching strategy, it is sub-optimal at least in two aspects: first, much better results would be expected if the program were delivered during the primary schools, when stereotypes are much less rooted in girls. Studies and research in psychology highlight how prejudices and stereotypes begin to form in the early years of life and become entrenched over time. Therefore, for an intervention that aims to be truly effective in shaping attitudes and beliefs, it would be appropriate to start working on these aspects at an even younger age, involving girls in inclusive and educational initiatives. An intervention at an older age thus results in greater difficulties and limitations, given that the number of girls interested in participating in projects of this type is initially limited by prejudices and constraints in self-perception. Biases and self-perception challenges can already be acting as barriers before such projects even begin. These pre-existing conditions can create a cycle that’s hard to break, where prejudices lead to low self-esteem and a lack of interest in opportunities that could actually help challenge and change those prejudices. Addressing these issues at an earlier stage and fostering a positive self-perception are crucial steps in encouraging broader participation and ensuring the success of such projects. Alternatively, to intervene at an older age, the support of teachers and families in encouraging girls, as well as strong media campaigns promoting the projects, would be important. Teachers and families play crucial roles in shaping perceptions and encouraging participation. Media campaigns can serve as powerful tools to raise awareness, change attitudes, and showcase the tangible benefits of involvement in such projects. Highlighting economic and career advantages could be particularly persuasive, offering clear and practical reasons for participation. Overall, a multi-faceted approach involving educational, familial, and media support is key to successfully engaging older participants and overcoming the initial barriers posed by prejudices and self-perception issues.

Another limitation is represented by the need to not stop at pilot initiatives, which often have a temporary duration and limited scope. The experiment of G4Greta involves only about 200 girls every year, which clearly limits the impact of its foreseen results. On the other side, scaling up requires, after two-three years of experimentation, the involvement of public bodies at the regional and/or national level. It is essential that projects are followed by structured and long-term interventions, capable of acting deeply and systematically on the mechanisms that underlie prejudices and stereotypes. Only in this way can we aspire to real and lasting change in girls’ participation to the Computer Science field. To implement long-term strategies and move beyond pilot projects, participation from decision-makers and the political sphere is necessary. In this perspective, it is important to involve key stakeholders in the process of creating and sustaining initiatives aimed at addressing biases and promoting equal opportunities. In this regard, G4Greta is already monitored by Repubblica Digitale, the national strategic initiative promoted by the Department for Digital Transformation of the Presidency of the Italian Council of Ministers. Decision-makers and political figures have the power to allocate resources, influence policies, and create an environment conducive to the success of such programs. Their commitment and active participation are crucial for ensuring that initial efforts are not just temporary but are instead followed by comprehensive and enduring strategies. This collaboration can help in establishing a solid foundation for these initiatives, providing them with the necessary support, and ensuring their long-term viability and impact.

Finally, a third limitation that we must face is that follow-up projects, necessary to monitor the effectiveness of initiatives over time and ensure continuity of action, require a significant financial commitment and a long time. Despite these challenges, it is essential to invest in these activities to ensure that the changes obtained are not ephemeral, but translate into a stable and positive transformation of social dynamics. Implementing follow-up procedures also ensures accountability and allows for the refinement of the programs based on empirical data and participant feedback. This iterative process of assessment and improvement is key to sustaining the relevance and effectiveness of the orientation programs, ensuring they continue to meet the evolving needs of the participants and remain aligned with the overarching goal of diversifying and enriching the field of Computer Science with more female talent.

In conclusion, the presented projects serve as a testament to the multifaceted benefits of embedding soft skills education within Computer Science orientation initiatives for girls. They not only illuminate the path toward achieving greater gender equity in the tech industry but also underscore the imperative for ongoing evaluation and adaptation of such programs to maximize their impact and efficacy.










\printcredits

\bibliographystyle{cas-model2-names}

\bibliography{cas-refs}

\begin{thebibliography}{18}
\expandafter\ifx\csname natexlab\endcsname\relax\def\natexlab#1{#1}\fi
\providecommand{\url}[1]{\texttt{#1}}
\providecommand{\href}[2]{#2}
\providecommand{\path}[1]{#1}
\providecommand{\DOIprefix}{doi:}
\providecommand{\ArXivprefix}{arXiv:}
\providecommand{\URLprefix}{URL: }
\providecommand{\Pubmedprefix}{pmid:}
\providecommand{\doi}[1]{\href{http://dx.doi.org/#1}{\path{#1}}}
\providecommand{\Pubmed}[1]{\href{pmid:#1}{\path{#1}}}
\providecommand{\bibinfo}[2]{#2}
\ifx\xfnm\relax \def\xfnm[#1]{\unskip,\space#1}\fi
\bibitem[{Alves et~al.(2014)Alves, Ribeiro and Machado}]{Alves}
\bibinfo{author}{Alves, L.}, \bibinfo{author}{Ribeiro, P.},
  \bibinfo{author}{Machado, R.}, \bibinfo{year}{2014}.
\newblock \bibinfo{title}{Project-based learning: An environment to prepare it
  students for an industry career}.
\newblock \bibinfo{journal}{Overcoming Challenges in Software Engineering
  Education: Delivering Non-Technical Knowledge and Skills} ,
  \bibinfo{pages}{230--249}.
\bibitem[{Berg et~al.(2018)Berg, Sharpe and Atkin}]{Berg}
\bibinfo{author}{Berg, T.}, \bibinfo{author}{Sharpe, A.},
  \bibinfo{author}{Atkin, E.}, \bibinfo{year}{2018}.
\newblock \bibinfo{title}{Females in computing: Understanding stereotypes
  through collaborative picturing}.
\newblock \bibinfo{journal}{Computers \& Education} \bibinfo{volume}{126},
  \bibinfo{pages}{105--114}.
\bibitem[{Bona et~al.(2023)Bona, Chanin, Nascimento and Sales}]{Bona23}
\bibinfo{author}{Bona, F.}, \bibinfo{author}{Chanin, R.},
  \bibinfo{author}{Nascimento, N.}, \bibinfo{author}{Sales, A.},
  \bibinfo{year}{2023}.
\newblock \bibinfo{title}{Understanding the gaps in software engineering
  education from the perspective of it leaders: A field study}.
\newblock \bibinfo{journal}{Proceedings of the 15th International Conference on
  Computer Supported Education} \bibinfo{volume}{2}, \bibinfo{pages}{511--518}.
\bibitem[{Casale(2020)}]{selfesteem}
\bibinfo{author}{Casale, S.}, \bibinfo{year}{2020}.
\newblock \bibinfo{title}{Gender Differences in Self-esteem and
  Self-confidence}. \bibinfo{publisher}{John Wiley \& Sons, Ltd}.
\newblock pp. \bibinfo{pages}{185--189}.
\newblock \URLprefix
  \url{https://onlinelibrary.wiley.com/doi/abs/10.1002/9781118970843.ch208},
  \DOIprefix\doi{https://doi.org/10.1002/9781118970843.ch208},
  \href{http://arxiv.org/abs/https://onlinelibrary.wiley.com/doi/pdf/10.1002/9781118970843.ch208}{\tt
  arXiv:https://onlinelibrary.wiley.com/doi/pdf/10.1002/9781118970843.ch208}.
\bibitem[{Cheryan et~al.(2013)Cheryan, Plaut, Hendron and Hudson}]{Cheryan}
\bibinfo{author}{Cheryan, S.}, \bibinfo{author}{Plaut, V.},
  \bibinfo{author}{Hendron, C.}, \bibinfo{author}{Hudson, L.},
  \bibinfo{year}{2013}.
\newblock \bibinfo{title}{The stereotypical computer scientist: Gendered media
  representations as a barrier to inclusion for women}.
\newblock \bibinfo{journal}{Sex Roles} \bibinfo{volume}{69},
  \bibinfo{pages}{58--71}.
\bibitem[{Groeneveld et~al.(2019)Groeneveld, Vennekens and Aerts}]{Groeneveld}
\bibinfo{author}{Groeneveld, W.}, \bibinfo{author}{Vennekens, J.},
  \bibinfo{author}{Aerts, K.}, \bibinfo{year}{2019}.
\newblock \bibinfo{title}{Software engineering education beyond the technical}.
\newblock \bibinfo{journal}{Proceedings of the 47th SEFI Conference} ,
  \bibinfo{pages}{1607--1622}.
\bibitem[{Jaccheri et~al.(2021)Jaccheri, Pereira and Fast}]{jaccheri2021gender}
\bibinfo{author}{Jaccheri, L.}, \bibinfo{author}{Pereira, C.},
  \bibinfo{author}{Fast, S.}, \bibinfo{year}{2021}.
\newblock \bibinfo{title}{Gender issues in computer science: Lessons learnt and
  reflections for the future}.
\newblock \href{http://arxiv.org/abs/2102.00188}{\tt arXiv:2102.00188}.
\bibitem[{Leonhardt and Overå(2021)}]{Leonhardt21}
\bibinfo{author}{Leonhardt, M.}, \bibinfo{author}{Overå, S.},
  \bibinfo{year}{2021}.
\newblock \bibinfo{title}{Are there differences in video gaming and use of
  social media among boys and girls? — a mixed methods approach}.
\newblock \bibinfo{journal}{Int J Environ Res Public Health}
  \bibinfo{volume}{18}, \bibinfo{pages}{6085}.
\bibitem[{Master et~al.(2021)Master, Meltzoff and Cheryan}]{Master}
\bibinfo{author}{Master, A.}, \bibinfo{author}{Meltzoff, A.},
  \bibinfo{author}{Cheryan, S.}, \bibinfo{year}{2021}.
\newblock \bibinfo{title}{Gender stereotypes about interests start early and
  cause gender disparities in computer science and engineering}.
\newblock \bibinfo{journal}{PNAS} \bibinfo{volume}{118,48}.
\bibitem[{O.D. Marcenaro-Gutierrez(2021)}]{Marcenaro}
\bibinfo{author}{O.D. Marcenaro-Gutierrez, L.A. Lopez-Agudo, C.H.},
  \bibinfo{year}{2021}.
\newblock \bibinfo{title}{Are soft skills conditioned by conflicting factors? a
  multiobjective programming approach to explore the trade-offs}.
\newblock \bibinfo{journal}{Economic Analysis and Policy} \bibinfo{volume}{72},
  \bibinfo{pages}{18--40}.
\bibitem[{Oguz(2019)}]{Oguz}
\bibinfo{author}{Oguz, D.;~Oguz, K.}, \bibinfo{year}{2019}.
\newblock \bibinfo{title}{Perspective on the gap between the software industry
  and the software engineering education}.
\newblock \bibinfo{journal}{IEEE Access} \bibinfo{volume}{7,1},
  \bibinfo{pages}{117527--117543}.
\bibitem[{Pang et~al.(2023)Pang, Li, Gao and Han}]{emphaty}
\bibinfo{author}{Pang, C.}, \bibinfo{author}{Li, W. nad~Zhou, Y.},
  \bibinfo{author}{Gao, T.}, \bibinfo{author}{Han, S.}, \bibinfo{year}{2023}.
\newblock \bibinfo{title}{Are women more empathetic than men? questionnaire and
  eeg estimations of sex/gender differences in empathic ability.}
\newblock \bibinfo{journal}{Soc Cogn Affect Neurosci.} \bibinfo{volume}{18:1},
  \bibinfo{pages}{511--518}.
\bibitem[{Rützler(2010)}]{Sutter}
\bibinfo{author}{Rützler, M.S.D.}, \bibinfo{year}{2010}.
\newblock \bibinfo{title}{Gender differences in competition emerge early in
  life}.
\newblock \bibinfo{journal}{IZA Discussion Paper} \bibinfo{volume}{5015},
  \bibinfo{pages}{1--33}.
\bibitem[{Saxa et~al.(2017)Saxa, Lehmana, Jacobsb, Kannyc, Limc, Monje-Paulsond
  and Zimmermanam}]{Saxa17}
\bibinfo{author}{Saxa, L.}, \bibinfo{author}{Lehmana, K.},
  \bibinfo{author}{Jacobsb, J.}, \bibinfo{author}{Kannyc, M.},
  \bibinfo{author}{Limc, G.}, \bibinfo{author}{Monje-Paulsond, L.},
  \bibinfo{author}{Zimmermanam, H.}, \bibinfo{year}{2017}.
\newblock \bibinfo{title}{Anatomy of an enduring gender gap: The evolution of
  women’s participation in computer science}.
\newblock \bibinfo{journal}{The Journal of Higher Education}
  \bibinfo{volume}{88, 2}, \bibinfo{pages}{258--293}.
\bibitem[{Sedelmaier(2015)}]{Sedelmaier}
\bibinfo{author}{Sedelmaier, Y;~Landes, D.}, \bibinfo{year}{2015}.
\newblock \bibinfo{title}{Swebos – the software engineering body of skills}.
\newblock \bibinfo{journal}{International Journal of Engineering Pedagogy}
  \bibinfo{volume}{5 (1)}, \bibinfo{pages}{12--19}.
\bibitem[{Seibel and N.(2019)}]{Seibel}
\bibinfo{author}{Seibel, S.}, \bibinfo{author}{N., V.}, \bibinfo{year}{2019}.
\newblock \bibinfo{title}{Factors influencing women entering the software
  development field through coding bootcamps vs. computer science bachelor’s
  degrees}.
\newblock \bibinfo{journal}{Journal of Computing Sciences in Colleges}
  \bibinfo{volume}{34, 6}, \bibinfo{pages}{84–96}.
\bibitem[{Siddiky(2020)}]{Siddiky}
\bibinfo{author}{Siddiky, M.R.}, \bibinfo{year}{2020}.
\newblock \bibinfo{title}{Does soft skill development vary among the students?
  a gender perspective}.
\newblock \bibinfo{journal}{Turkish Journal of Education} \bibinfo{volume}{9,
  3}, \bibinfo{pages}{205--221}.
\bibitem[{Vachovsky et~al.(2016)Vachovsky, Wu, Chaturapruek, Russakovsky,
  Sommer and Fei-Fei}]{Vachosky}
\bibinfo{author}{Vachovsky, M.}, \bibinfo{author}{Wu, G.},
  \bibinfo{author}{Chaturapruek, S.}, \bibinfo{author}{Russakovsky, O.},
  \bibinfo{author}{Sommer, R.}, \bibinfo{author}{Fei-Fei, L.},
  \bibinfo{year}{2016}.
\newblock \bibinfo{title}{Towards more gender diversity in cs through an
  artificial intelligence summer program for high school girls}.
\newblock \bibinfo{journal}{Proceedings of the 47th ACM Technical Symposium on
  Computing Science Education} , \bibinfo{pages}{303--308}.

\end{thebibliography}

\bio{}
\endbio

\endbio

\end{document}